\documentclass[a4paper,usenatbib]{article}

\usepackage{epsfig}
\usepackage{graphicx}

\begin{document}

\label{firstpage}

\begin{center}

\LARGE{\bf A novel type of intermittency in a nonlinear dynamo in a compressible flow}\\

\vspace{0.5cm}

\Large{Erico L. Rempel,$^{1,2}$ Michael R. E. Proctor$^{1}$ and \\
Abraham C.--L. Chian$^{3}$}\\
\end{center}

\vspace{0.5cm}

\noindent \small{$^{1}$Department of Applied Mathematics and Theoretical Physics (DAMTP), University of Cambridge, Cambridge CB3 0WA, UK\\
\noindent $^{2}$Institute of Aeronautical Technology (IEFM/ITA) and World Institute
for Space Environment Research (WISER), S\~ao Jos\'e dos Campos -- SP 12228--900,
Brazil\\
\noindent $^{3}$National Institute for Space Research (INPE) and World Institute
for Space Environment Research (WISER), P. O. Box 515,
S\~ao Jos\'e dos Campos -- SP 12227--010, Brazil}

\vspace{0.5cm}

\begin{center}
\normalsize{\bf Abstract}

\end{center}

{\small The transition to intermittent mean--field dynamos is studied using
numerical simulations of isotropic magnetohydrodynamic turbulence driven by a helical flow.
The low-Prandtl number regime is investigated by keeping the kinematic viscosity 
fixed while the magnetic diffusivity is varied.
Just below the critical parameter value for the onset of dynamo action, a transient 
mean--field with low magnetic
energy is observed. After the transition to a sustained dynamo,
the system is shown to evolve through different types of intermittency
until a large--scale coherent field with small--scale turbulent fluctuations
is formed. Prior to this coherent field stage, a new type of intermittency is 
detected, where the magnetic field randomly alternates between phases
of coherent and incoherent large--scale spatial structures. 
The relevance of these findings to the understanding of the physics of mean--field 
dynamo and the physical mechanisms behind intermittent behavior observed in  
stellar magnetic field variability are discussed.}

\begin{center}
{\bf keywords:}
magnetic fields -- MHD -- turbulence.
\end{center}

\section{Introduction}

	The observation of strong magnetic fields in astrophysical bodies 
(planets, stars, and galaxies) suggests the existence of a dynamo process, whereby a weak 
(seed) magnetic field is amplified due to the conversion of kinetic energy into magnetic
 energy. Dynamos can be classified as large-scale (or mean--field) dynamos, or small--scale 
(or fluctuation) dynamos, according to whether the magnetic fields grow on spatial scales 
larger or smaller than the energy carrying scale of the fluid motion. 
A typical manifestation of a large--scale dynamo is the solar cycle, where the distribution
of sunspots in space and time display large--scale spatial coherence and long--term temporal correlation,
as seen in the {\it butterfly diagram} \cite{solanki03,solanki06,thomasweiss08}. 
Although the maxima and minima of solar activity
form a recurrent 11--year cycle, long periods of very low solar activity, grand minima such as the Maunder minimum
\cite{beer98},
have led several authors to look for a description of the solar cycle as
an intermittent event due to the chaotic nature of the dynamo, driving the system 
to random alternations between phases of ``regular magnetic activity'' and grand minima.
Many works employ low--dimensional models based on ordinary differential equations to investigate these modulations
[e.g., Covas et al. \cite{covas97} and Wilmot-Smith et al. \cite{wilmot05}].
Others have focused on high--dimensional mean--field models based
on partial differential equations.
Covas \& Tavakol \cite{covas99} reported the existence of crisis-induced and Pomeau--Manneville type--I
intermittencies \cite{manneville79,chian98,chian06,chian_etal07}
in an axisymmetric mean--field dynamo model, suggesting the {\it multiple--intermittency hypothesis},
since more than one type of intermittency may be responsible for the minima observed in solar cycle data.
Ossendrijver \cite{ossendrijver00} showed that a 2--D mean-field dynamo model that features an $\alpha$--effect based on 
the buoyancy instability of magnetic flux tubes could produce grand minima.
The grand minima observed by this model were later described by Ossendrijver \& Covas \cite{ossendrijver03} as a 
manifestation of crisis--induced intermittency \cite{grebogi83,rempel05,rempel07,rempel04c,rempel_etal:2007}.
Intermittency in a 2--D mean-field model was also reported by Moss \& Brooke \cite{moss00},
where the reaction of the Lorenz force on the rotation is used as the nonlinear effect that
limits the magnetic field at finite amplitude.
Charbonneau \cite{charbonneau04} showed that an axisymmetric 2--D solar cycle model based on the 
Babcock-Leighton mechanism of poloidal
field regeneration can exhibit intermittency in the presence of low--amplitude noise.
Recently, Brandenburg \& Spiegel \cite{brandenburg08} observed on--off intermittency in a mean--field dynamo
model after imposing stochastic fluctuations in the $\alpha$--effect or using a fluctuating
electromotive force. A review on the use of intermittent chaotic models to capture the 
main qualitative aspects of the temporal and spatial variability of the 
solar--cycle is provided by Spiegel \cite{spiegel09}.

This paper employs 3--D magnetohydrodynamics (MHD) simulations to investigate the onset of intermittent mean--field 
dynamo as a function of the magnetic Prandtl number $Pr$,
defined as the ratio between the kinematic viscosity $\nu$ and the magnetic diffusivity $\eta$. Although the model geometry is highly idealised, the full induction and momentum equations are solved, so that there is no need to appeal to an averaging process.
Early numerical works in similar models focused on $Pr \geq 1$ regimes due to difficulties in exciting
a nonlinear dynamo for low values of Pr [see, e.g., Nordlund et al. \cite{nordlund92}]. Nevertheless, this regime is crucial for understanding
astrophysical plasmas, where $Pr$ is usually much less than one [$Pr \sim 10^{-7}$--$10^{-4}$ in the
solar convective zone, depending on the depth \cite{schekochihin07}]. Furthermore,
the role of compressibility of the velocity field in astrophysical plasmas should not be 
ignored, even in low--Mach--number regimes \cite{kleeorin97,haugen04}. Thus, we 
adopt a compressible MHD code and explore low--$Pr$ regimes.
Two conditions 
have been shown to be important for a fluid flow to act as a large--scale dynamo: chaotic 
stream--lines and kinetic helicity \cite{childress95,biskamp00}. 
Depending on the characteristics of the velocity field and other parameters, such as $\eta$,
the resulting magnetic field lines may 
display regular or irregular motion, or a decay to a purely hydrodynamic 
state. 
In order to obtain a chaotic and helical velocity field, the 
flow is driven by an ABC (Arnold--Beltrami--Childress) forcing, which is a superposition of three
helical waves \cite{arnold65,galloway84} with a characteristic wave number $k_f$,
which sets the energy carrying scale of the flow. The emergence of a mean field 
is studied by forcing the flow at scale $k_f=5$ and observing the energy transfer
towards larger scales.

This paper is divided as follows. Section \ref{sec2} contains a description of the dynamo model.
The main results are described in section \ref{sec3}.
The kinetic viscosity is set at a value such that the velocity field is
 weakly turbulent. Then, a seed magnetic field is applied and the magnetic diffusivity $\eta$ 
is progressively reduced until the onset of dynamo is observed. Three types of dynamo regimes 
are observed as a function of $\eta$. In the first regime, right after the onset of dynamo, there is 
an intermittent switching between bursty phases of high--amplitude magnetic activity and
quiescent phases of low magnetic energy, similar to the on--off spatiotemporal intermittency 
reported in other works \cite{sweet01a,sweet01b,rempel07,rempel_etal:2007}. 
The second regime reveals an intermittent switching between coherent and incoherent large--scale
structures and, up to our knowledge, has not been reported in previous dynamo studies.
This intermittent process persists up to a certain threshold value of the magnetic diffusivity, 
where the third regime is observed, with the system self--organizing into a spatially coherent mean--field 
that exhibits complex temporal dynamics. An analysis of the spatiotemporal complexity 
of the patterns observed in each  regime is also described.
The conclusions are given in section \ref{sec conc}.

\section[]{THE MODEL} \label{sec2}

We consider a compressible isothermal gas (the ratio of specific heats $\gamma = c_p/c_v = 1$, 
where $c_p$ is the heat capacity at constant pressure and $c_v$ is the heat capacity at constant volume) 
with constant sound speed $c_s$, constant dynamical viscosity $\mu$, constant magnetic diffusivity 
$\eta$, and constant magnetic permeability $\mu_0$. 
The continuity equation is solved in terms of the logarithm of the density $\ln \rho$, since this quantity
varies spatially much less than density.

\begin{equation}
\frac{\partial\ln\rho}{\partial t}+\mathbf{u}\nabla\ln\rho+\nabla\cdot\mathbf{u}=0,
\label{eq continuity}
\end{equation}
where $\mathbf{u}$ is the fluid velocity.
The momentum equation is given by

\begin{equation}
\frac{\partial\mathbf{u}}{\partial t}+\mathbf{u\cdot}\nabla\mathbf{u}=-\frac{\nabla p}{\rho} +\frac{\mathbf{J\times B}}{\rho}+\frac{\mu}{\rho}\left(\nabla^{2}\mathbf{u}+\frac{1}{3}\nabla\nabla\cdot\mathbf{u}\right)+\mathbf{f},
\label{eq momentum}
\end{equation}
where $\mathbf{J} = \nabla \times \mathbf{B}/\mu_0 $ is the current density, $p$ is the pressure and  
$\mathbf{f}$ is an external forcing. The pressure gradient is obtained from the entropy equation for an ideal gas 
$s=c_v \ln(p\rho^{-\gamma})$. Thus, $\nabla p / \rho = c_s^2\ln\rho$, 
where  $c_s^2 = \gamma p/\rho$ is assumed to be constant. 
The induction equation is written in terms of the magnetic vector potential $\mathbf{A}$, so that 
$\nabla \cdot \mathbf{B} = 0$, since $\mathbf{B} = \nabla \times \mathbf{A}$

\begin{equation}
\frac{\partial\mathbf{A}}{\partial t}=\mathbf{u\times B}-\eta\mu_{0}\mathbf{J}.
\label{eq induction}
\end{equation}
We adopt nondimensional units, 
such that $c_s = \rho_0 = \mu_0 = 1$, where $\rho_0=\left<\rho\right>$ is the spatial average of $\rho$. 
Equations (\ref{eq continuity})--(\ref{eq induction}) are solved with the 
PENCIL CODE\footnote{\tt http://www.nordita.org/software/pencil-code} in a box with 
sides $L = 2\pi$ and periodic boundary conditions. The PENCIL CODE is a compressible MHD code that
employs high-order finite differences and has been extensively used in
astrophysical simulations [see Brandenburg \& Subramanian \cite{brandenburg05} and references therein].
The initial condition is $\ln \rho = \mathbf{u} = 0$, 
and $A$ is a set of normally distributed, uncorrelated random numbers with zero mean
and standard deviation 
equal to $10^{-3}$. For the forcing function $\mathbf{f}$ we adopt the form of ABC flow used in 
Sur et al. \cite{sur07} 

\begin{equation}
\mathbf{f(x)}=\frac{A_{f}}{\sqrt{3}}\left(\begin{array}{c}
\sin k_{f}z+\cos k_{f}y\\
\sin k_{f}x+\cos k_{f}z\\
\sin k_{f}y+\cos k_{f}x\end{array}\right),
\end{equation}
where $A_f$ is the amplitude and $k_f$ the wavenumber of the forcing function, which is isotropic
with respect to the three coordinate directions.
The ABC forcing is an interesting choice for dynamo studies, since it is a Beltrami flow, 
$\nabla \mathbf{u} \propto \mathbf{u}$, 
and, therefore, is maximally helical, i.e., 
$(H_k)^2 = \left<\mathbf{u}\cdot\nabla\times\mathbf{u}\right>^2 = \left<|\mathbf{u}|^2\right>\left<|\nabla\times\mathbf{u}|^2\right>$, 
where the angle brackets 
denote spatial integration over the periodic domain. 

In all the following sections we use $A_f = 0.1$, which yields Mach numbers below $0.5$.
A numerical resolution of $64^3$ mesh points is chosen. As in Brandenburg \cite{brandenburg01},
we set $k_f = 5$ in order to 
be able to see the emergence of a large scale magnetic 
field, with spatial scales larger than the energy injection scale. 
Spatial averages are denoted by $\left< \cdot \right>$ and 
time averages by $\left< \cdot \right>_t$. Unless otherwise stated, 
references to kinetic $(Re)$ and magnetic $(Rm)$ Reynolds numbers
are based on the forcing scale 

\begin{equation}
Re = \frac{\lambda_fU}{\nu},\qquad Rm = \frac{\lambda_fU}{\eta}, 
\end{equation}
where $\nu = \mu / \rho_0$ is the average kinematic viscosity,
$\lambda_f=2\pi/k_f$ is the forcing spatial scale, and $U=\left<u^2\right>^{1/2}$ is the mean velocity
at a time when the magnetic field is saturated.

\section{RESULTS} \label{sec3}

\subsection{Hydrodynamic simulations}

\begin{figure}
 \includegraphics[width=1.\columnwidth]{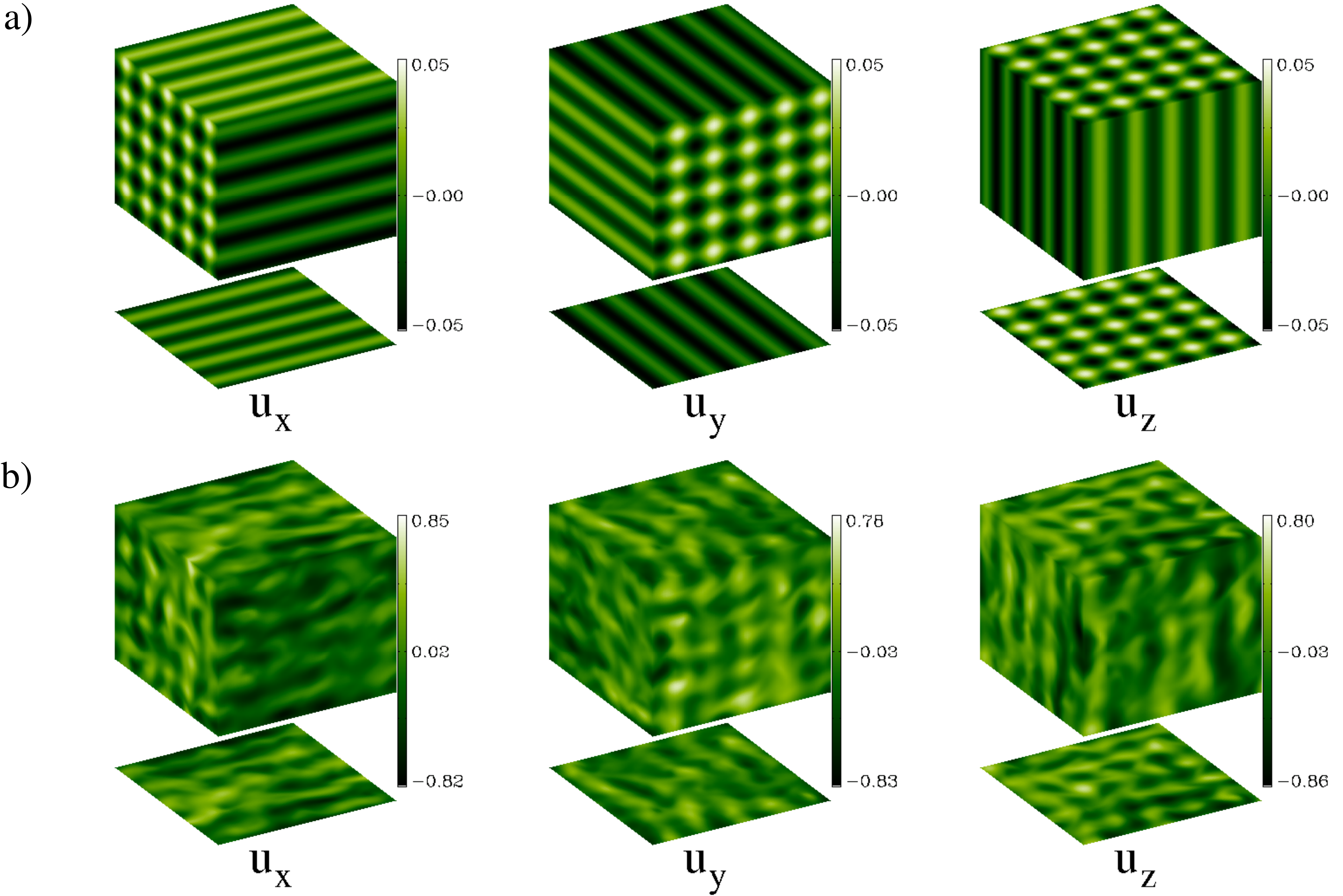}
 \caption{(Colour online) Contour plots of the velocity components for hydrodynamic 
simulations at (a) $\nu=0.02$ and (b) $\nu=0.005$.}
\label{fig1}
\end{figure}

\begin{figure}
\begin{center}
 \includegraphics[width=.5\columnwidth]{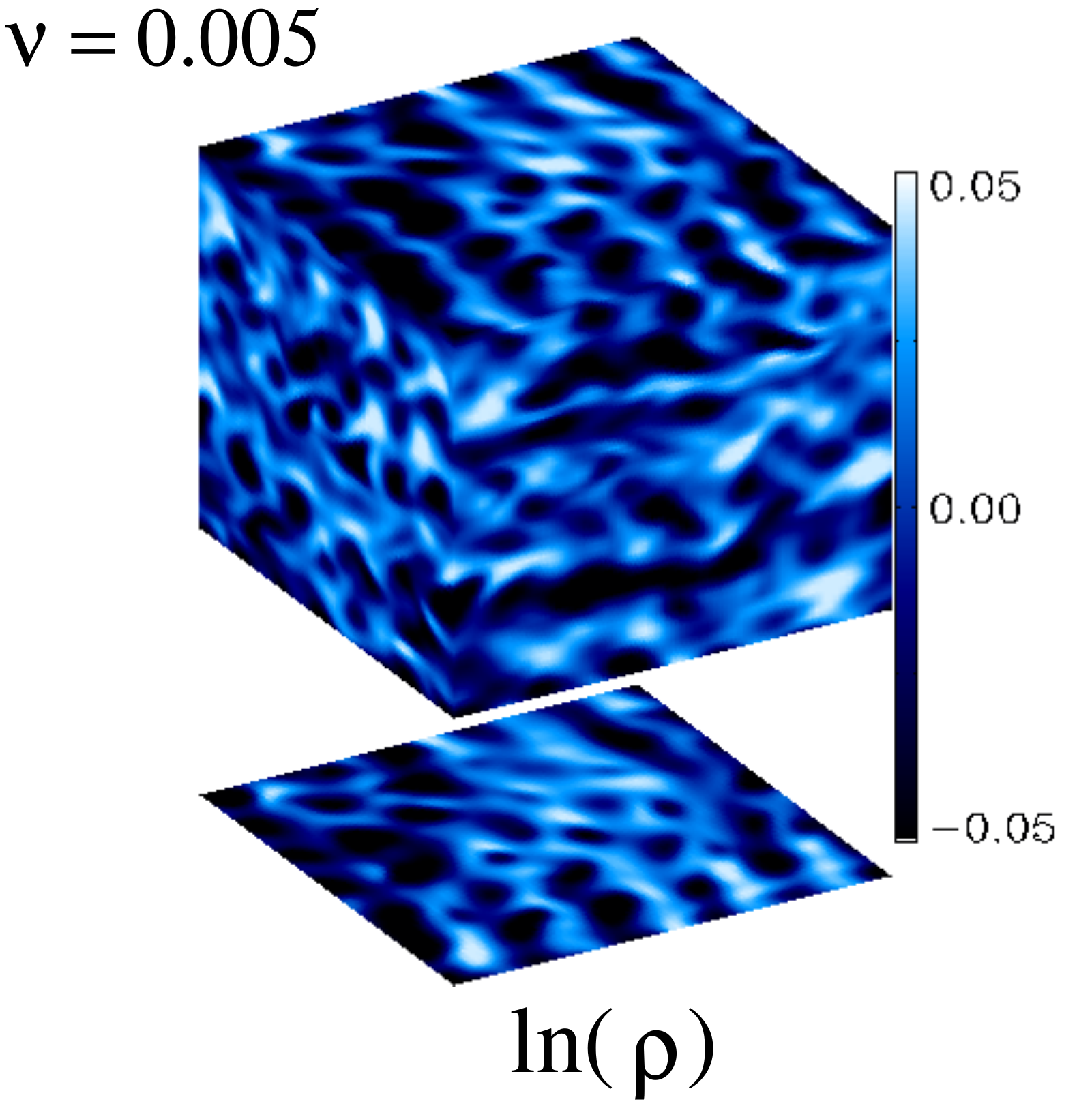}
\end{center}
 \caption{(Colour online) Contour plot of ln($\rho$) for the hydrodynamic simulation at $\nu=0.005$.}
\label{fig2}
\end{figure}

We start with hydrodynamic simulations in the absence of magnetic fields. 
We set $k_f=5$, $A=0.1$, and vary the average kinematic viscosity $\nu$. 
Figure \ref{fig1}(a) displays the velocity components of an asymptotically stable solution of
Eqs. (\ref{eq continuity})--(\ref{eq momentum}) at
$\nu = 0.02$, which corresponds to $Re \sim 12.38$. The initial velocity field ${\bf u=0}$
evolves with time until it takes the shape of the ABC flow.
After converging to this attractor, the amplitudes of the velocity 
components do not change with time, so this is a steady state of the system.
As the kinematic viscosity is reduced, the ABC flow becomes unstable and a sequence of 
symmetry breaking bifurcations takes place.
At $\nu = 0.005$ ($Re \sim 100$ based on the forcing scale, or $Re \sim 500$ based on the box scale), 
the system is in a weakly turbulent regime, 
as illustrated in Fig. \ref{fig1}(b). 
The flow is weakly compressible, as can be seen in Fig. \ref{fig2}, which
shows that the density fluctuations are within $5 \%$ of the mean value $\rho_0=1$.

\subsection{Onset of dynamo action}

\begin{figure}
\begin{center}
 \includegraphics[width=1.\columnwidth]{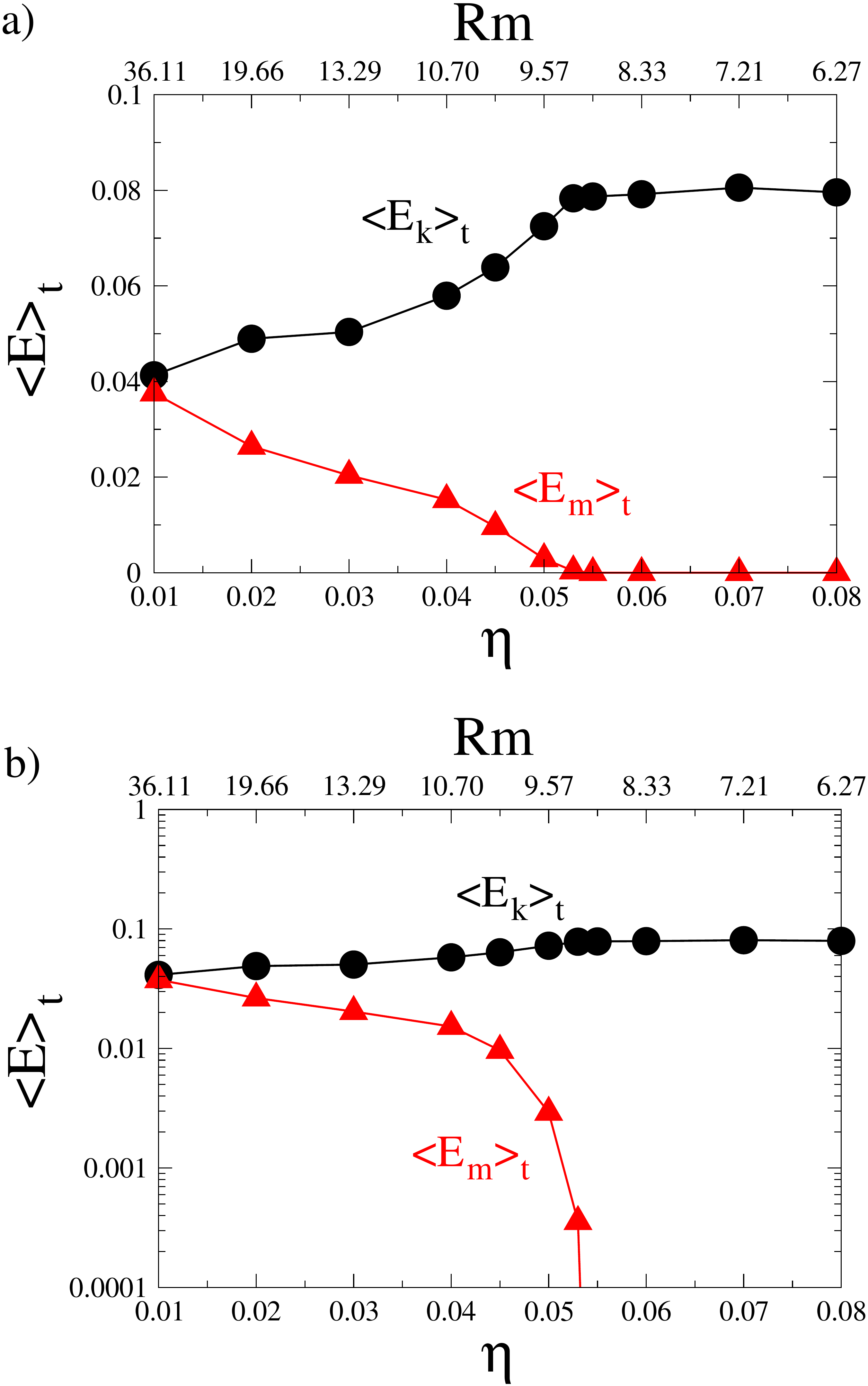}
\end{center}
 \caption{(Colour online) Variation of the time--averaged kinetic (black circles) and magnetic (red triangles)
energies as a function of $\eta$ in (a) linear and (b) linear--log scales. The upper scales display the 
magnetic Reynolds number based on the forcing scale.}
\label{fig3}
\end{figure}

We now fix $\nu = 0.005$ 
and change the magnetic diffusivity $\eta$ as we look for the onset 
of nonlinear dynamo action. Since we are interested in the low--Prandtl number limit, 
$\eta$ is made larger than $\nu$, as $Pr$ is varied from $0.0625$ to $0.5$. 
Figure \ref{fig3} shows the bifurcation diagrams for the time--averaged magnetic ($\left<E_m\right>_t$, red triangles) 
and kinetic ($\left<E_k\right>_t$, black circles) energies in linear [Fig. \ref{fig3}(a)] and 
linear--log [Fig. \ref{fig3}(b)] 
scales as a function of $\eta$ (lower axes) or $Rm$ (upper axes). 
For each value of $\eta$, the initial variations of $\left<B^2\right>/2$ 
and $\left<u^2\right>/2$ are discarded, 
and time--averages are
computed for the saturated energy values. For large values of $\eta$, the seed magnetic field 
decays rapidly and there is no dynamo. 
At the  onset of dynamo action at  $\eta \sim 0.053$ ($Rm \sim 9.5$ in the forcing scale, or $Rm \sim 47.5$ in the box scale), 
the magnetic energy starts to grow at the 
expense of kinetic energy. 
Two different dynamo behaviours can be identified in Fig. \ref{fig3}. In the first range, between $\eta \sim 0.053$ and 
$\eta \sim 0.04$ there is a sharp increase in the saturated magnetic energy.
Between $\eta \sim 0.04$ and $\eta \sim 0.01$ the magnetic energy increases more slowly as an
exponential function of $\eta$ until it is comparable to the kinetic energy at $\eta=0.01$.

\begin{figure}
\begin{center}
 \includegraphics[width=1.\columnwidth]{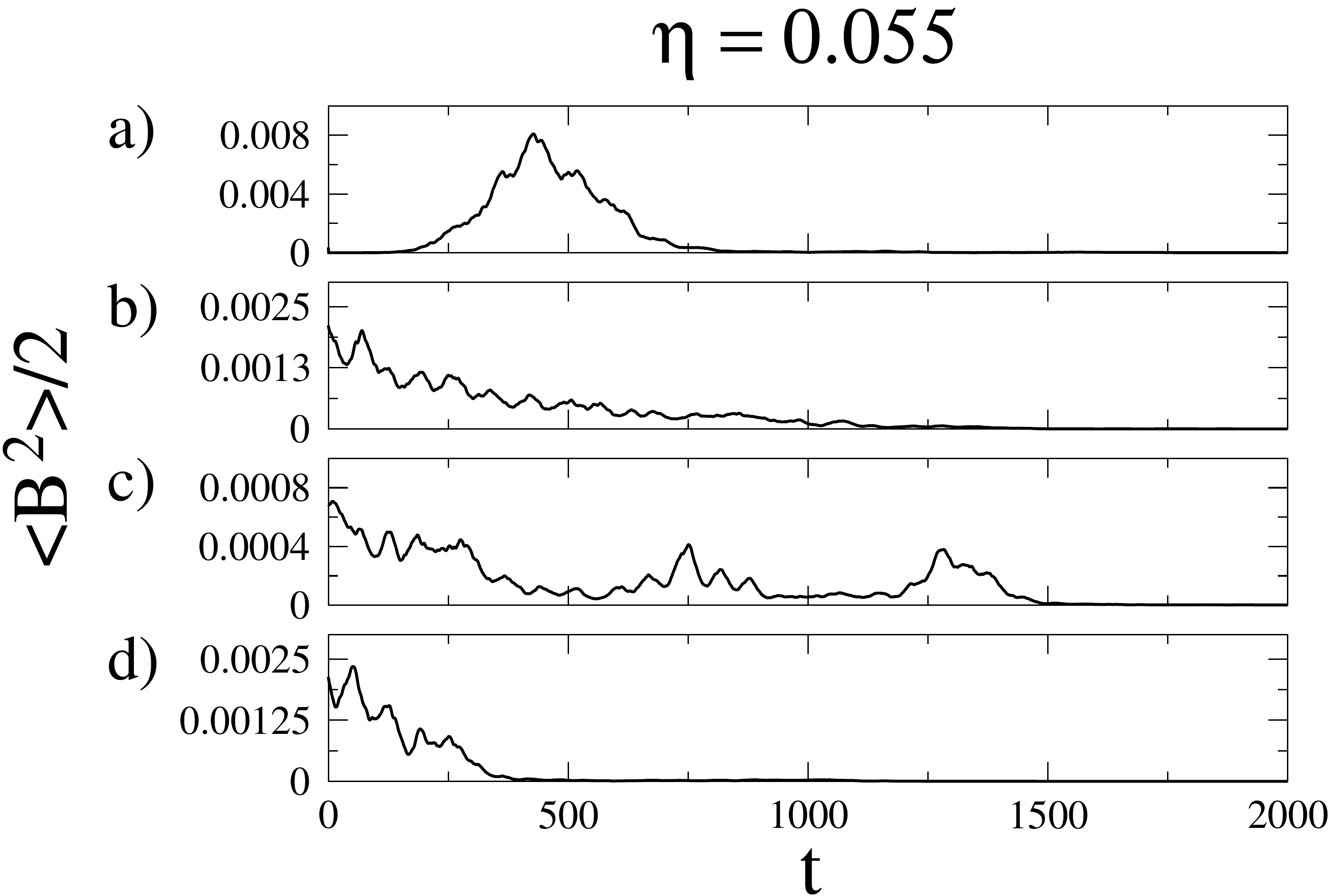}
\end{center}
 \caption{Time series of magnetic energy at $\eta=0.055$ exhibiting transient dynamics for different initial conditions.}
\label{fig4}
\end{figure}

We turn to the dynamics near the transition to nonlinear dynamo.
Figure \ref{fig4} plots the time series of magnetic energy for different initial conditions at 
$\eta=0.055$,
just before the onset of dynamo action.
For a small seed magnetic field there is an initial exponential growth of the magnetic energy [Fig. \ref{fig4}(a)], 
as expected from kinematic dynamo theory. 
Then the field saturates and starts to decay. 
Even during the growth phase the magnetic field has a much lower magnitude than 
the velocity field. Consequently, the impact of the Lorentz force 
on the velocity field is negligible and the same
kind of behaviour was observed in kinematic simulations, where the Lorentz force term $\mathbf{J\times B}/\rho$
was removed from Eq. (\ref{eq momentum}). Figures \ref{fig4}(b)--\ref{fig4}(d) show other instances 
of a ``transient dynamo'' for different types of initial conditions. Thus, the dynamo state
is not an attractor of the system and we believe this transient and apparently chaotic dynamics displayed by the 
magnetic field is a signature of nonattracting chaotic sets known as
{\it chaotic saddles} \cite{hsu88,rempel_etal:2004,rempel04b,chian_etal07}, 
which have attracted wide attention recently due to their role  
in transition to hydrodynamic turbulence  \cite{peixinho06,hof06,willis07,hof08}.  

\subsection{Intermittent dynamo} \label{sec31}

\begin{figure}
\begin{center}
 \includegraphics[width=1.\columnwidth]{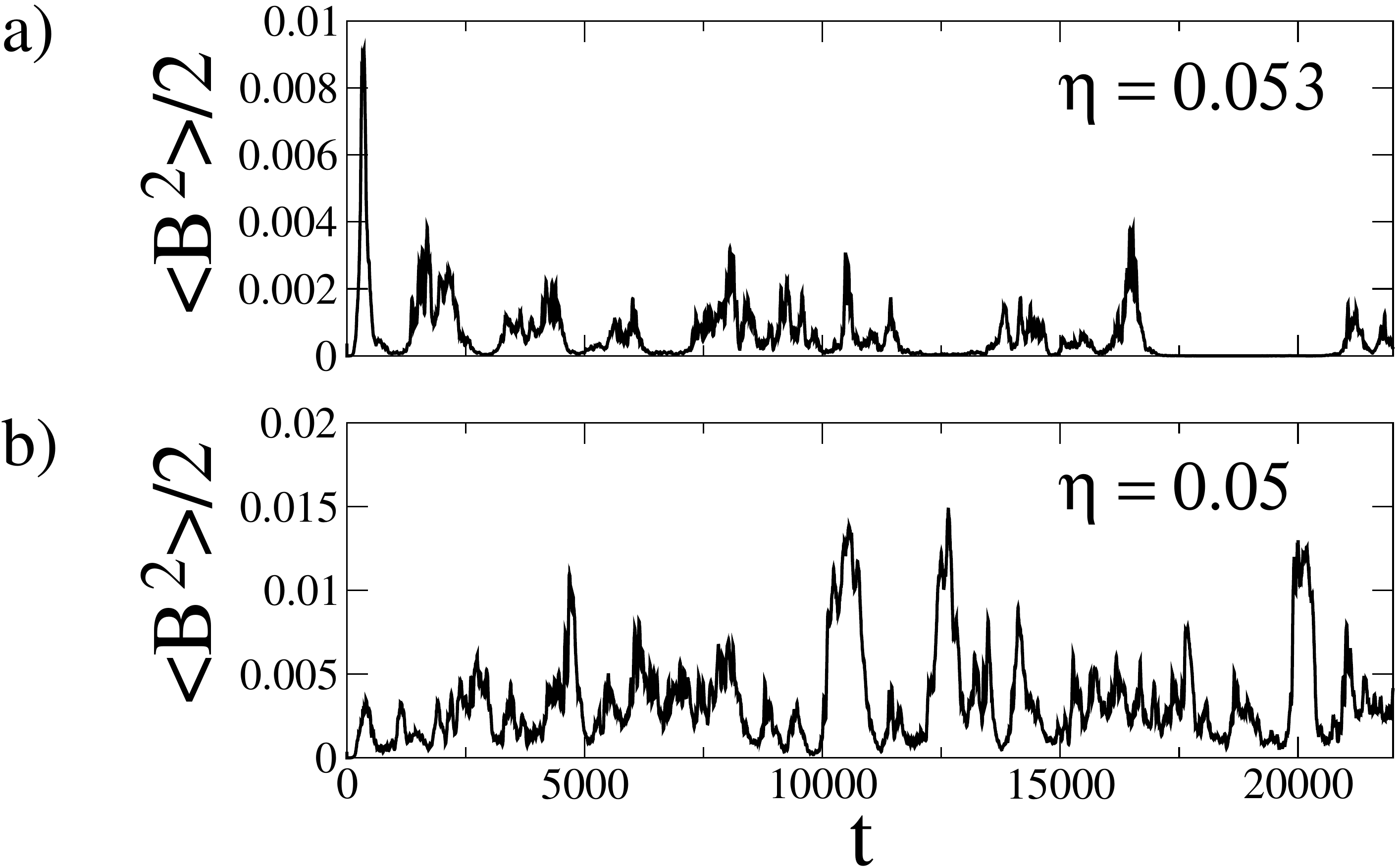}
\end{center}
 \caption{Intermittent time series of the magnetic energy at (a) $\eta=0.053$ and (b) $\eta=0.05$.}
\label{fig5}
\end{figure}

At $\eta = 0.053$, there is a transition to sustained dynamo action. 
The time series of magnetic energy exhibit random switching between phases of bursty 
and quiescent activity, as seen in Fig. \ref{fig5}(a), including long periods of minima with almost zero magnetic activity
such as the one between $t \sim 17500$ and $t \sim 20000$. 
This is the same type of behaviour reported by Sweet et al. \cite{sweet01a,sweet01b} for low Reynolds number ($Re=6.3$) dynamo
simulations with Prandtl number close to one and ABC-forcing at scale $k_f=1$. A blowout bifurcation, whereby
the hydrodynamic state looses transversal stability, was characterized as responsible for the intermittency,
 which in this case is called {\it on--off intermittency}, since the solutions
go arbitrarily close to a manifold defined by the purely hydrodynamic state $\mathbf{B=0}$
(``on'' phases) and suddenly depart from the manifold during the strong magnetic bursts (``off'' phases).

\begin{figure}
\begin{center}
 \includegraphics[width=1.\columnwidth]{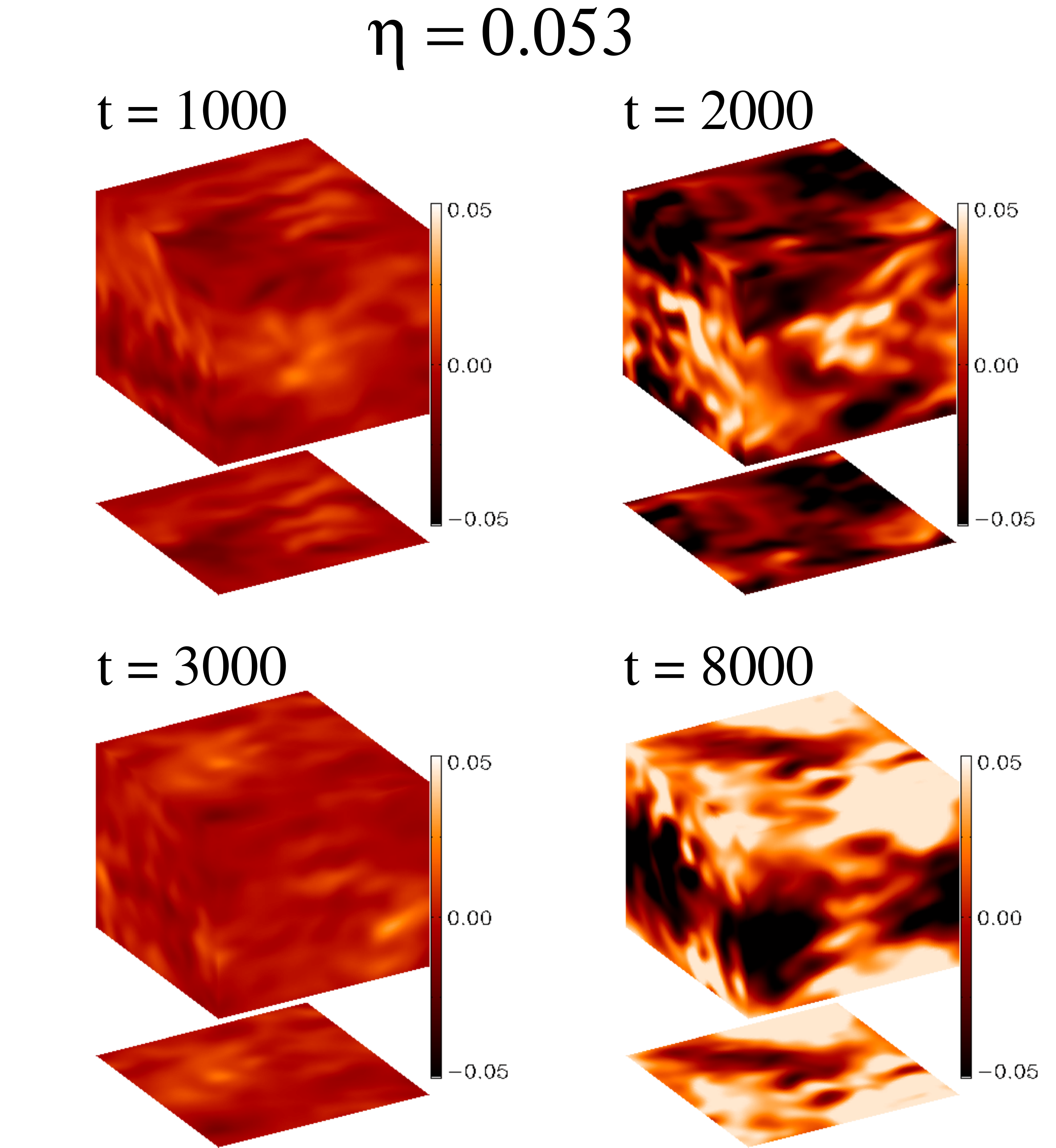}
\end{center}
 \caption{(Colour online) Contour plots of $B_x$ for $\eta=0.053$ at four different values of $t$. 
The magnetic energy increases and
decreases randomly in time, as in Fig. \ref{fig5}(a), resulting in on--off intermittency.}
\label{fig6}
\end{figure}

\begin{figure}
\begin{center}
 \includegraphics[width=1.\columnwidth]{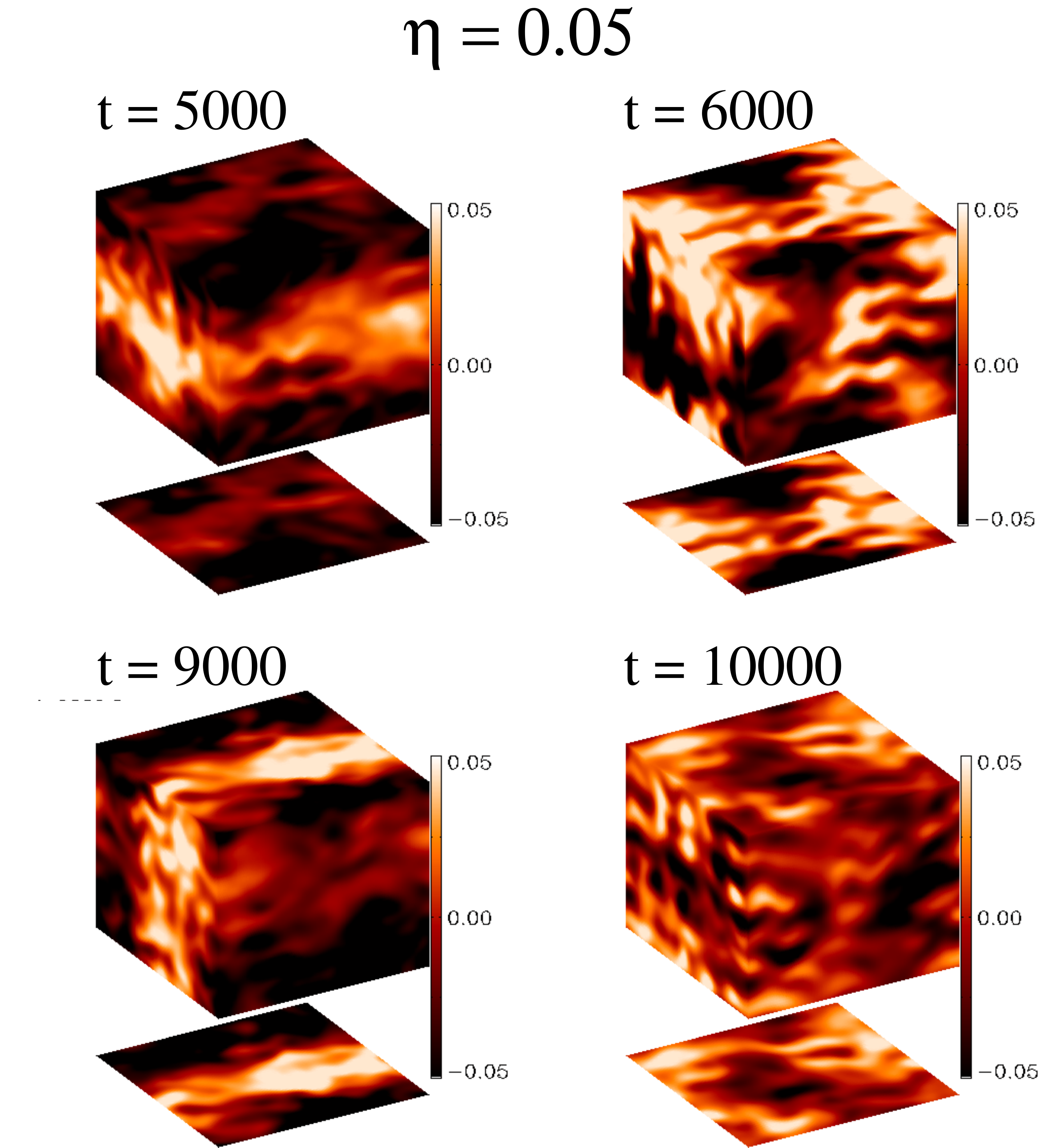}
\end{center}
 \caption{(Colour online) Contour plots of $B_x$ for $\eta=0.05$ at four different values of $t$. There is an
intermittent switching between large scale approximately sinusoidal states 
(at $t=5000$ and $t=9000$, for instance) and incoherent states.}
\label{fig7}
\end{figure}

For  $\eta = 0.05$, the dynamo is strongly intermittent, but no grand minima are found 
in the time series of $\left<B^2\right>/2$, shown in Fig. \ref{fig5}(b). There are strong bursts
interspersed by lower peaks, but the energy is seldom close to zero. A look at the spatial
structures of the magnetic field is helpful to distinguish the two types of intermittency
exemplified by Figs. \ref{fig5}(a) and \ref{fig5}(b).
Figure \ref{fig6} shows the contour plots of $B_x$ for the on--off intermittency ($\eta=0.053$) at five different 
times. The snapshots in the left column (predominantly red) are taken at quiescent phases, 
and the ones in the right column, at bursty phases. The magnetic field displays complex spatial 
structures in both phases, although they are more clearly seen in the bursty phases.
The dynamics at $\eta=0.05$ is very different. The contour plots of $B_x$, shown in Fig. \ref{fig7}, 
reveal that a large 
scale sinusoidal modulation of the magnetic field is evolving, but there is an intermittent 
switching in the preferred direction selected by the magnetic field. In the first snapshot $(t = 5000)$ 
the preferred direction for $B_x$ is $z$; in the 
second snapshot $(t = 6000)$ the field is restructuring and there is no preferred direction; in the third 
snapshot $(t = 9000)$ the preferred direction is $y$ and there is no preferred direction at $t=10000$.
This coherent--incoherent intermittency has been observed for all tested values of $\eta$ in the interval
from $\eta=0.05$ to $\eta = 0.02$.

\begin{figure}
\begin{center}
 \includegraphics[width=1.\columnwidth]{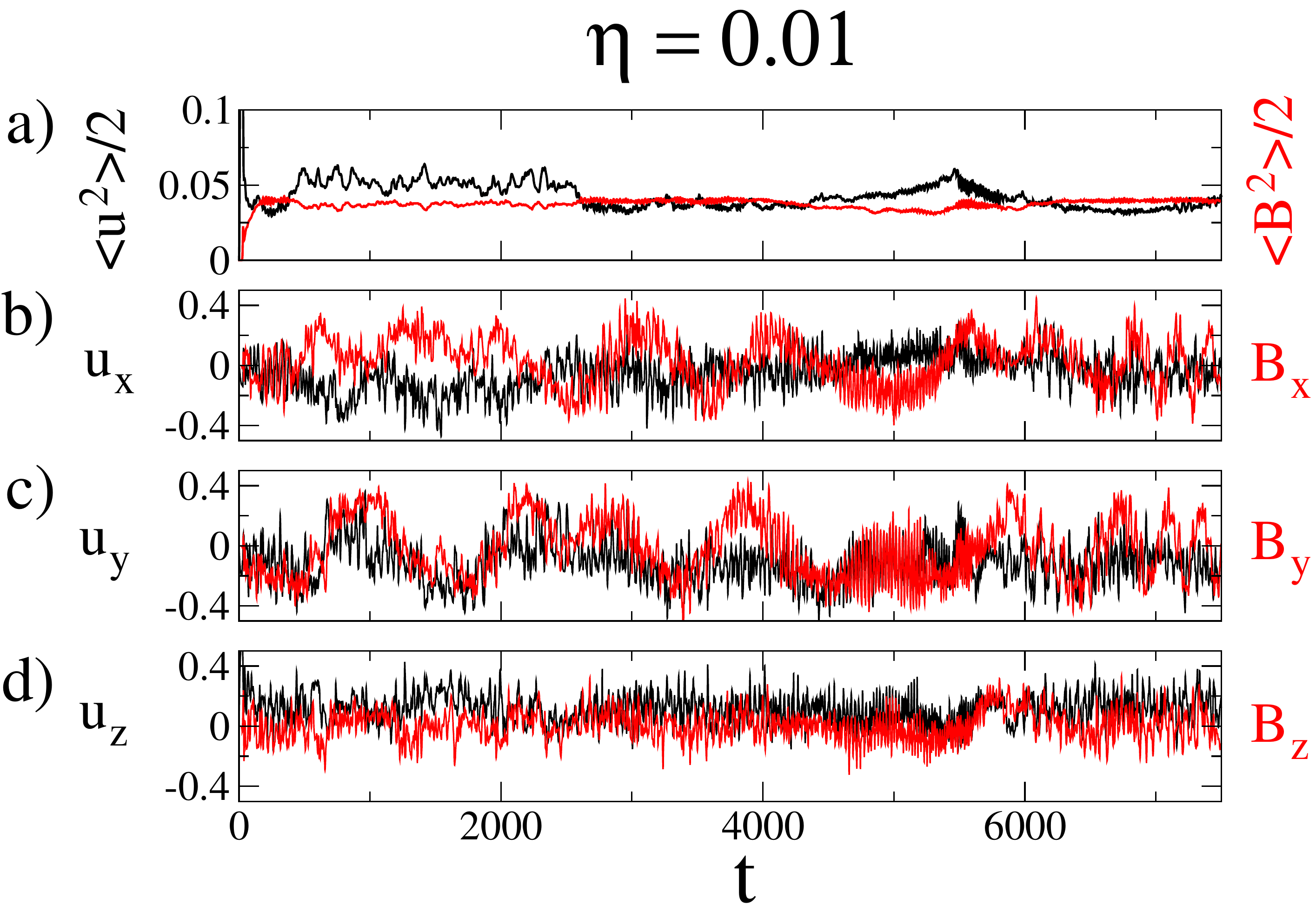}
\end{center}
 \caption{(Colour online) (a) Time series of the magnetic (red) and kinetic (black) energies at $\eta=0.01$;
(b)--(d) time series of the magnetic (red) and velocity (black) field components at a given
point in the $64^3$ numerical mesh.}
\label{fig8}
\end{figure}

\begin{figure}
\begin{center}
 \includegraphics[width=1.\columnwidth]{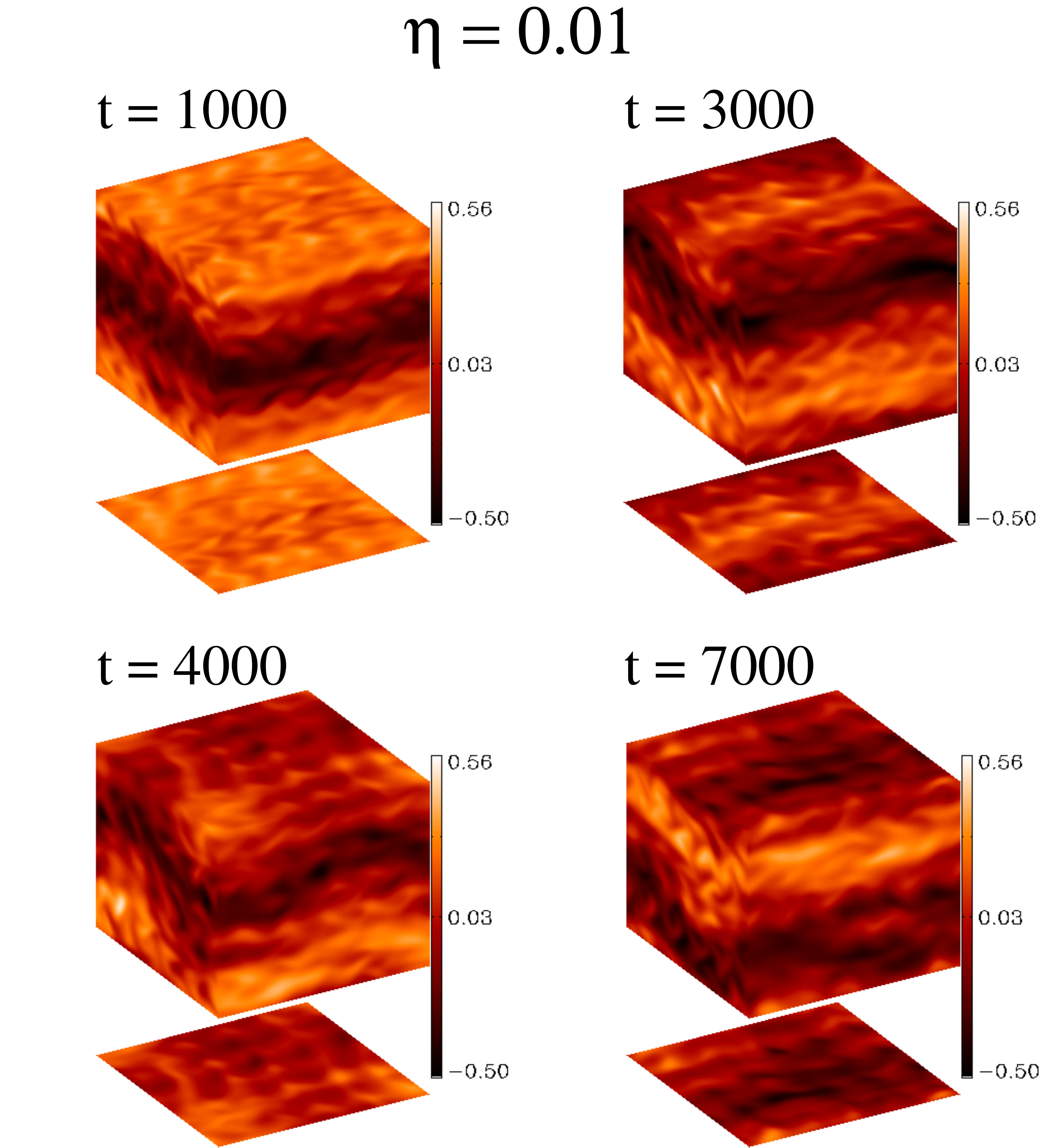}
\end{center}
 \caption{(Colour online) Contour plots of $B_x$ for $\eta=0.01$ at four different values of $t$.
A robust large--scale coherent structure can be observed throughout the simulation.}
\label{fig9}
\end{figure}

For  $\eta = 0.01$ the magnetic field has settled to a large scale and approximately sinusoidal mean field, 
with complex small scale spatial structures and irregular oscillations, as shown in Figs. \ref{fig8} and 
\ref{fig9}. In Fig. \ref{fig8}(a) the magnetic energy $\left<B^2\right>/2$ is strong enough to 
cause a reduction of the mean kinetic energy from about $\left<u^2\right>/2=0.4$ in the previous examples, 
to $\left<u^2\right>/2= 0.3$.  
There is also a definite correlation between the time series of the velocity and magnetic field components.

\subsection{Inverse cascade}

\begin{figure}
\begin{center}
 \includegraphics[width=1.\columnwidth]{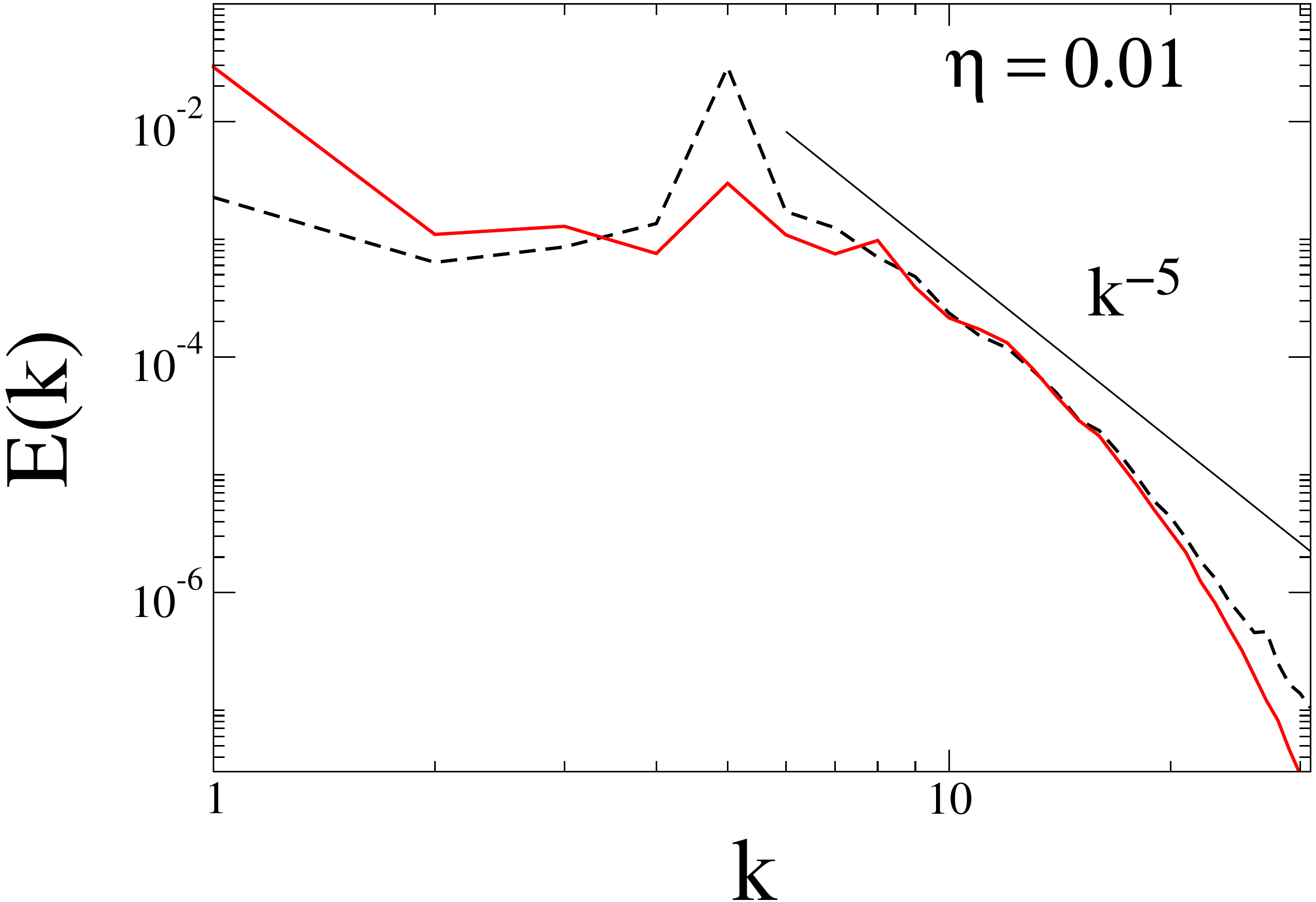}
\end{center}
 \caption{(Colour online) Magnetic (solid red line) and kinetic (dashed black line) energy spectra 
at $t=5000$ for $\eta=0.01$. The $k^{-5}$ line is a guide to the eye.}
\label{fig10}
\end{figure}

The appearance of a mean--field dynamo is due to the 
transfer of magnetic energy from the energy injection scale to larger scales.
This transfer is expected in helical flows and in mean-field dynamo theory is
attributed to the $\alpha$--effect 
\cite{moffatt78,krauseradler80}, whereby
interactions between  low-scale fluctuations of $\mathbf{B}$ and $\mathbf{u}$
produce a large--scale $\mathbf{B}$ [other mechanisms can be responsible for 
the rise of a mean-field in the absence of a mean $\alpha$--effect, such as the interaction of a 
fluctuating $\alpha$--effect and large-scale shear flows 
\cite{proctor07,hughes09}.
The kinetic and magnetic energy power spectra for the 
coherent mean--field state at $\eta = 0.01$ are 
shown in Fig. \ref{fig10} for $t=5000$. The one--dimensional spectra are obtained from the three--dimensional
spectra by computing the integrals
 of the spectral energy along spherical shells with rays defined by the modulus of $\mathbf{k}$.
For the magnetic spectrum (solid red line), although the forcing scale is $k_f = 5$, there is a 
backward transfer of 
magnetic energy, making $k = 1$ the predominant scale, resulting in a $\mathbf{B}$ field with a 
typical scale of the size of the box. As for the kinetic energy (dashed black line), 
$k = 5$ is still the predominant scale.
For this value of $Rm$ the spectra are particularly similar in the inertial range (not properly defined yet, 
since they are not sufficiently extended).
The $k^{-5}$ line
is a guide to the eye, and is the power--law observed for the magnetic spectrum of dynamos with
helical flows at low $Rm$ (following a $k^{-3}$ range) \cite{muller04,mininni07}.

\begin{figure}
\begin{center}
 \includegraphics[width=1.\columnwidth]{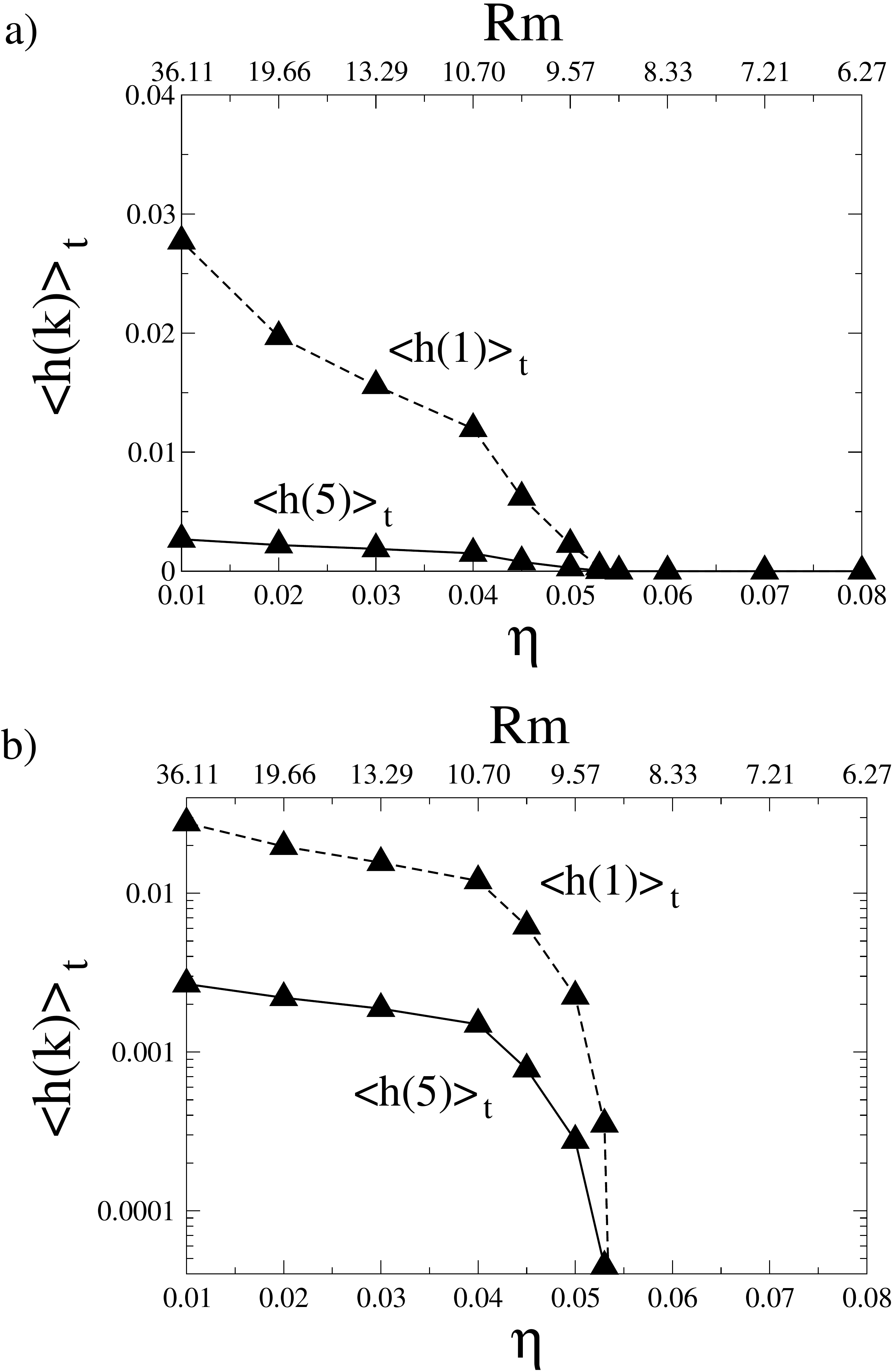}
\end{center}
 \caption{Peak height of modes $k_1$ (dashed lines) and $k_5$ (solid lines) of the 
time--averaged magnetic energy spectra as a function of $\eta$ in (a) linear and (b) linear--log scales.}
\label{fig11}
\end{figure}

Figure \ref{fig11} plots the peak height for the Fourier 
modes $k = 1$ (dashed line) and $k = 5$ (solid line) of the time--averaged magnetic energy spectra
 as a function of $\eta$.
It shows that the energy difference between these scales increases as the magnetic diffusivity decreases,
giving rise to a progressively more coherent mean--field, as discussed in the next section. 
The two distinct ranges previously identified
in Fig. \ref{fig3} can be seen in Fig. \ref{fig11} as well. The first
one with a steepest energy increase between $\eta =0.053$ and $\eta=0.04$, and the second one
between $\eta=0.04$ and $\eta=0.01$, where energy seems to increase exponentially, as seen
in the linear--log plot of Fig. \ref{fig11}(b).

\subsection{Spatiotemporal complexity} \label{sec complex}

Two operators computed in the Fourier space can be used to quantify the spatiotemporal complexity 
of a system. 
The amount of spatial disorder can be quantified by means of the 
spectral entropy \cite{powell_percival:1979,xi_gunton:1995,rempel_etal:2007}
\begin{equation} \label{eq_spectral_entropy}
  S(t) = - \sum_{k = 1}^{M} p_{k}(t) \ln(p_{k}(t)),
\end{equation}
where $M$ is the number of modes and $p_{k}(t)$ is the relative weight of a Fourier mode $b_k$ at an instant $t$
\begin{equation} \label{eq_relative_weight}
  p_{k}(t) = \frac{\left| b_k(t) \right|^2}{\sum_{k=1}^M \left| b_k(t) \right|^2}.
\end{equation}
Since $\mathbf{u}$ and $\mathbf{B}$ are real variables, 
\begin{eqnarray}
   |b_{-k}(t)|  & = &  |b_k(t)|, \label{eq_realvars_LW}
\end{eqnarray}
and only Fourier modes with $k > 0$ need to be considered.  
Mode $b_0(t)$ is decoupled from the other modes and is null for all $t$ if $b_0(0)=0$. 
The spectral entropy is maximum for a random system with uniform distribution, 
i.e., for all $k$, $p_{k}(t) = 1/M$, in which case $S(t) = \ln M$ 
\cite{badii_politi:1997}. For the present spatial resolution, $M=32$ and
the maximum entropy is $S_{\mathrm{MAX}} \sim 3.47$.

Another quantity of interest is the average wave number, or spectral average
\cite{thyagaraja79,lopes99,he03}

\begin{equation}
N(t)=\frac{\sqrt{\sum_{k=1}^{M}k^2\left|b_k(t)\right|^2}}{\sqrt{\sum_{k=1}^M\left|b_k(t)\right|^2}}.
\end{equation}
The spectral average is a measure of the number of active modes. In the absence of energy cascade,
$N(t)$ is limited by the number of linearly unstable modes. 

\begin{figure}
\begin{center}
 \includegraphics[width=1.\columnwidth]{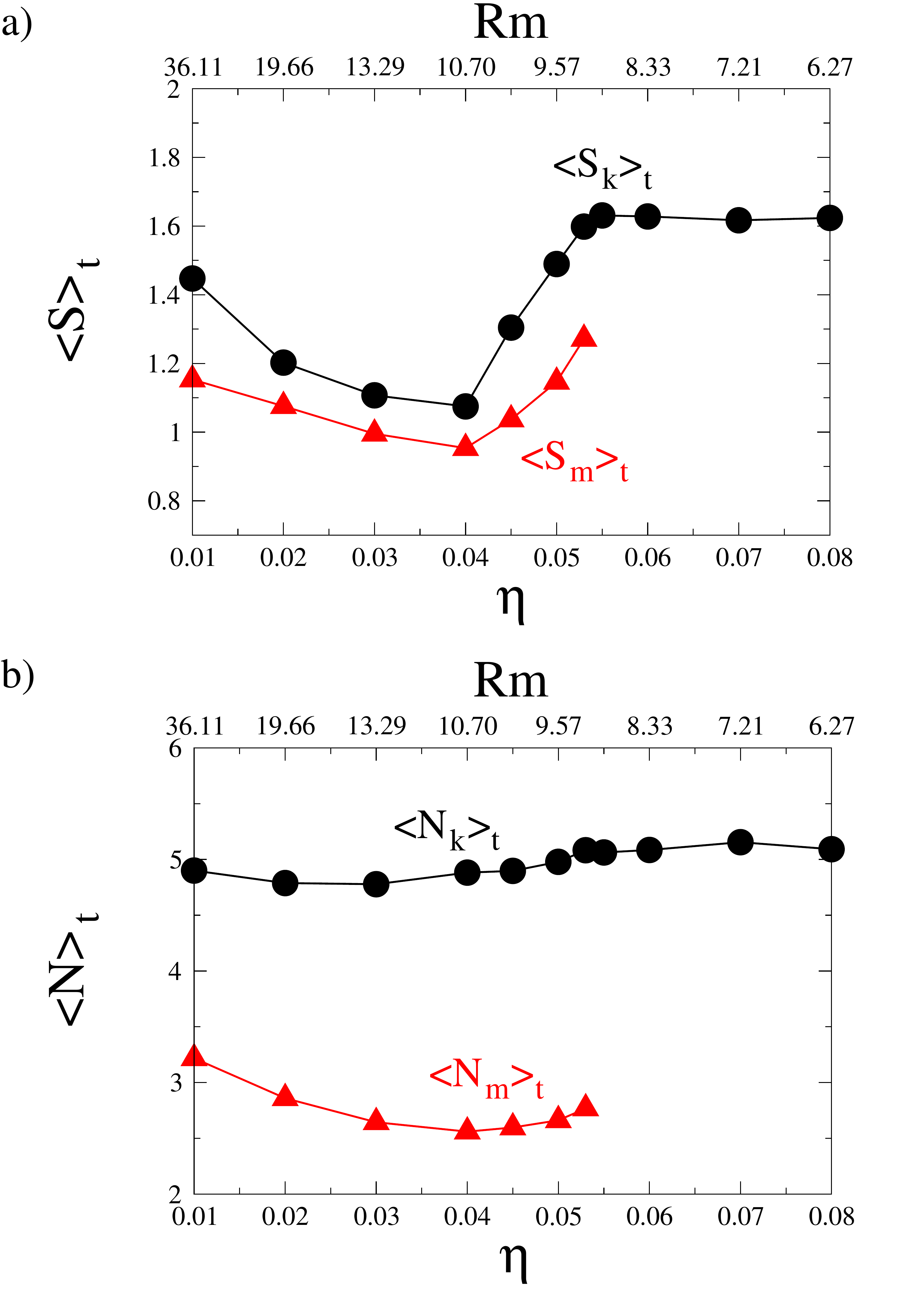}
\end{center}
 \caption{(Colour online) (a) Time--averaged spectral entropy as a function of $\eta$
for the kinetic (black circles) and magnetic (red triangles) energy spectra; 
(b) time--averaged spectral average as a function of $\eta$ for the kinetic (black circles)
and magnetic (red triangles) energy spectra.}
\label{fig12}
\end{figure}

Figure \ref{fig12} shows the time--averaged spectral entropy $\left<S\right>_t$ (top panel) and the 
time--averaged spectral average $\left<N\right>_t$ (bottom panel) 
for the kinetic (black circles) and magnetic (red triangles) energy spectra as a function of $\eta$. 
It can be concluded from the top panel that the spatial structures of the velocity field become 
less complex with the onset 
of the on--off intermittent dynamo at $\eta \sim 0.053$,
with a decrease in $\left<S_k\right>_t$. This is due to the beginning of the 
action of the Lorenz force on the momentum equation.
From $\eta =0.055$ to $\eta=0.04$ there is a strong decay not only in $\left<S_k\right>_t$, 
but also in $\left<S_m\right>_t$,  
reflecting the increase in the frequency of occurrence of large--scale 
coherent mean--field structures in the coherent--incoherent intermittent dynamo as $\eta$ is decreased.
The spatial complexity of both the velocity and magnetic fields, then, starts to increase from 
$\eta \sim 0.04$ to $\eta \sim 0.01$. This tendency coincides with an increase in the magnetic 
spectral average $\left<N_m\right>_t$ (bottom panel), which reflects a rise in magnetic energy in smaller scales. 
Note that the kinetic spectral average $\left<N_k\right>_t$ is always close to the energy injection scale
$k=5$, and the magnetic spectral average is lower due to the high energy peak in $k=1$ 
(see Figs. \ref{fig10} and \ref{fig11}).

\begin{figure}
\begin{center}
 \includegraphics[width=1.\columnwidth]{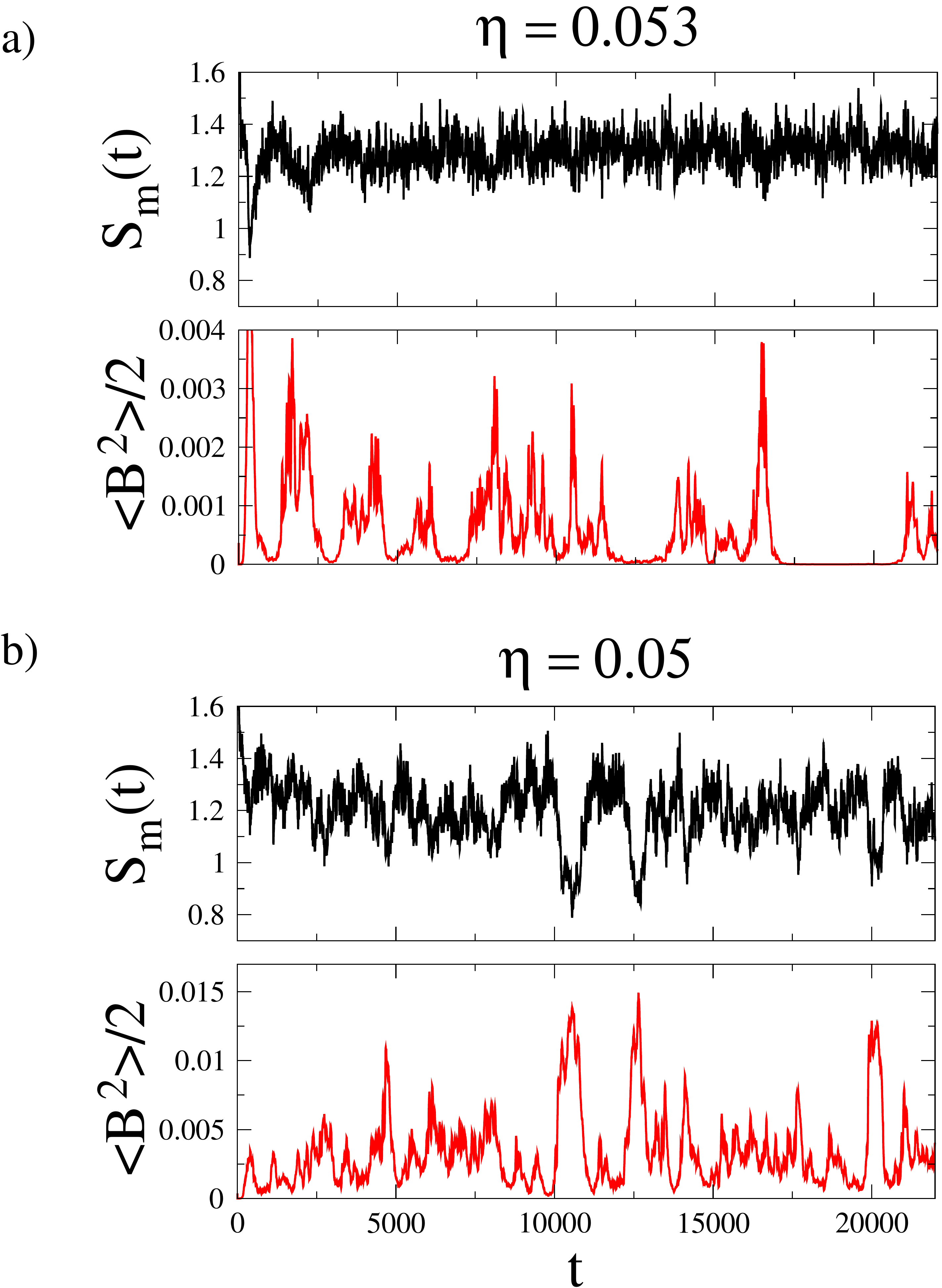}
\end{center}
 \caption{(Colour online) (a) Time series of the spectral entropy $S_m(t)$ 
and magnetic energy $\left<B^2\right>/2$ for the on--off intermittency at $\eta=0.053$;
(b) same as (a) but for the coherent--incoherent intermittency
at $\eta=0.05$.}
\label{fig13}
\end{figure}

The spectral entropy $S_m(t)$ is able to distinguish the two types of intermittency
displayed in Fig. \ref{fig5} of section \ref{sec31}. Figure \ref{fig13}(a) 
shows the time series of the spectral entropy $S_m(t)$ 
and magnetic energy $\left<B^2\right>/2$ for the on--off intermittency at $\eta=0.053$. 
There is little correlation between the two series,
since $S_m(t)$ oscillates erratically even during phases of grand minima in the magnetic energy time
series. 
Figure \ref{fig13}(b) plots the same time series as Fig. \ref{fig13}(a),
but for the coherent--incoherent intermittency at $\eta=0.05$. There is now strong
correlation between $S_m(t)$ and $\left<B^2\right>/2$, as the entropy decreases whenever 
the magnetic energy increases abruptly during spatially coherent bursts, indicating
a high--variability not only in the energy amplitude, but also in the spatial complexity.

\section{Conclusions} \label{sec conc}

We have described an intermittent route to mean--field dynamos
as a function of the (low) magnetic Prandtl number $Pr$. 
The main contributions
of this work are the quantification of spatial complexity 
of the magnetic field patterns from 3--D compressible MHD dynamo simulations 
(Figs. \ref{fig12} and \ref{fig13}), and
the identification of a new type of intermittency characterized by
the switching between coherent and incoherent large-scale structures 
[Figs. \ref{fig5}(b), \ref{fig7}, and \ref{fig13}(b)].
The mechanisms responsible for
this intermittency have not yet been explored  
and, in future works, we expect that a better description 
can be achieved by means of the analysis of phase synchronization 
of Fourier modes \cite{koga08,chian09}.

This paper is also an addition to a series of previous works on the onset 
of nonlinear dynamo action in ABC flows.
Kinematic and nonlinear dynamos were studied by Galanti et al. \cite{galanti92} for several choices of 
the forcing scale $k_f$, 
and $Pr$ ranging from 1 to 13, with low kinetic Reynolds number $Re$ (up to 20); 
A bifurcation study with low-$Re$, $Pr=1$, and $k_f=1$ was conducted by Seehafer et al. \cite{seehafer96};
Sweet et al. \cite{sweet01a,sweet01b} detected a blowout bifurcation responsible for on--off intermittency
for $k_f=1$, $Re=6.3$ and $Pr$ close to one;
Brummell et al. \cite{brummell01} studied kinematic and nonlinear regimes with $Re$ between 50 and 100 and $k_f=1$
in a time--dependent ABC flow previously introduced by Galloway \& Proctor \cite{galloway92};
Mininni \cite{mininni07} investigated the inverse energy cascade at small $Pr$ (down to 0.005) and $k_f=3$,
with $Re$ varying from 11 to 6200; Alexakis \& Ponty
\cite{alexakis08} studied the effect of the Lorenz force on on--off intermittency in ABC flows
with $k_f=1$ and varying both $Re$ and $Rm$.
In contrast to all the aforementioned papers which performed incompressible
MHD simulations, in this paper we opted for a compressible code. It is not yet clear whether the compressibility is a crucial feature of the dynamics observed, though certainly compressibility is a feature shared by the solar dynamo!
Our simulations were performed with $k_f=5$,
$Re \sim 100$ (or $Re \sim 500$ for the box-scale $Re$) and $Pr \in \left[0.0625, 0.5\right]$.
Despite the use of a model with simple geometry, reasonably small kinetic
and magnetic Reynolds numbers, and
without differential rotation, our results reveal some qualitative
resemblance to certain aspects of the solar dynamo and we expect them to be useful in the 
analysis of more realistic models.

\noindent{\bf Acknowledgments}

ELR thanks DAMTP for their kind hospitality.
This work has been supported by CNPq (Brazil) and FAPESP (Brazil).

\label{lastpage}

\end{document}